# Thermal infrared emission reveals the Dirac point movement in biased graphene

Marcus Freitag, Hsin-Ying Chiu, Mathias Steiner, Vasili Perebeinos, Phaedon Avouris

IBM Thomas J. Watson Research Center, Yorktown Heights, New York 10598, USA

**Graphene is a 2-dimensional material with high carrier mobility[1, 2] and thermal conductivity,[3, 4] suitable for high-speed electronics. Conduction and valence bands touch at the Dirac point. The absorptivity of single-layer graphene is $\alpha=2.3\%$, nearly independent of wavelength.[5, 6] Here we investigate the thermal radiation from biased graphene transistors. We find that the emission spectrum of single-layer graphene follows that of a grey body with constant emissivity $\varepsilon=(1.6\pm0.8)\%$. Most importantly, we can extract the temperature distribution in the ambipolar graphene channel, as confirmed by Stokes/anti-Stokes measurements. The biased graphene exhibits a temperature maximum whose location can be controlled by the gate voltage. We show that this peak in temperature reveals the spatial location of the minimum in carrier density, i.e. the Dirac point.**

To explore the graphene thermal radiation, we fabricated large, rectangular transistors from exfoliated graphene, with lithographically-defined source and drain contacts and a backgate to control the carrier concentration. When a current is passed through the graphene, electrical energy is transformed into Joule heat, which is mainly dissipated into the substrate and the metallic contacts.[7] Some of the energy is radiated into free space, which we detect spatially or spectrally resolved in the photon energy range between 0.5 eV to 0.95 eV. Details of the experimental setup (Fig. 1a) and sample fabrication can be found in the Methods section. Figure 1c shows three infrared spectra, acquired for electrical power densities between 5.5 kW cm$^{-2}$ and 18.6 kW cm$^{-2}$. We fit a total of 11 spectra to the formula for a grey body (Planck's law, modified by an emissivity $\varepsilon$),

$$I(\nu,T) = \varepsilon \frac{8\pi}{h^2 c^3} \frac{(h\nu)^3}{\exp(h\nu/k_b T - 1)} \; , \quad (1)$$

where $h\nu$ is the photon energy and $T$ is the temperature. Corresponding drain voltages and currents are shown in Fig. 1d. The prefactor $\varepsilon \frac{8\pi}{h^2 c^3}$ is first determined from the spectra corresponding to the highest electrical power densities. It is then fixed at that value for all the blackbody fits to determine the various temperatures, which range from 390 K at 5.5 kW cm$^{-2}$ to 530 K at 18.6 kW cm$^{-2}$ (Fig. 1e). The good fit quality in Fig. 1c suggests that graphene indeed behaves like a grey body with constant emissivity and deviations are of the order of ±20%, a value that is likely limited by uncertainties in the optical system response rather than the graphene emissivity itself.

After calibration with a blackbody source of known temperature and emissivity, we extracted an emissivity value of ε=(1.6±0.8)%. This value is in good agreement with the measured absorptivity of α=2.3% for single-layer graphene. [5] It is important to note that our graphene samples are electrically biased and therefore could deviate from thermal equilibrium. Nevertheless we found that Kirchhoff's law holds within experimental errors in the measured photon energy range of 0.5 to 0.95 eV. The total radiative loss can be estimated using the Stefan-Bolzmann law $I = \varepsilon \sigma A T^4$, where $\sigma = 5.67 \cdot 10^{-8}$ Wm$^{-2}$K$^{-4}$ is the Stefan-Bolzmann constant and $A = (55 \times 3.5)$ μm$^2$ is the area of the graphene sheet. For a drain bias $V_d$=-40 V and temperatures around $T$=530 K, the radiative loss is 20 nW, less than $10^{-6}$ of the total power. The major part of the electrical power is dissipated non-radiatively into the substrate.

The validity of the temperature measurement through thermal emission was confirmed using another sample where we employed two established methods to extract the local temperature (see Fig. 2). (I) Raman Stokes/anti-Stokes measurements that determine temperature by measuring the phonon occupation number $n$ of the G-phonon band:

$$\frac{I_S}{I_{AS}} = \frac{1+n}{n}\left(\frac{E_L - \hbar\omega}{E_L + \hbar\omega}\right)^3 \left(\frac{\chi_S}{\chi_{AS}}\right)^2 \approx \exp\left(\frac{\hbar\omega}{kT}\right)\left(\frac{E_L - \hbar\omega}{E_L + \hbar\omega}\right)^3. \qquad (2)$$

In eq. 2, $\hbar\omega$ is the energy of the G phonon, $E_L$ is the laser excitation energy, and $\chi_{S, AS}$ are Stokes/anti-Stokes Raman susceptibilities. Since the graphene absorption is essentially constant, both $\chi_S$ and $\chi_{AS}$ are the same, and the measurements can be performed at the same $E_L$. (II) The shift of the Raman G' band that scales with the temperature as $\Delta\omega_{G'}/\Delta T = -0.034$ cm$^{-1}$/K. [8] Both Raman methods show good agreement with the temperatures extracted from the thermal radiation.

Recently, non-equilibrium phonon distributions have been found in electrically-heated carbon nanotubes.[9-13] In that case, the optical phonons G and G', as well as intermediate-frequency phonons, are populated at much higher "temperatures" than the radial-breathing mode (RBM) and other acoustic phonons. In graphene, there is no RBM, and this proxy for the acoustic phonon temperature is not available for Raman measurements. The infrared emission however, represents an average temperature of all the phonons. The close agreement of this temperature with the two Raman-derived temperatures suggests that no significant non-equilibrium phonon distribution exists between acoustic and optical phonons in graphene at least up to 700K. A non-equilibrium G-band phonon population was recently observed in a graphene constriction. [14] There, the temperature derived from G-band Stokes/anti-Stokes measurements was much larger (1500K) than the temperature derived from the G-band red shift (500K). The strong local electric fields in the constriction were sufficient to drive the G-band phonon population out of equilibrium. The authors also showed that the constriction emits visible light that can be detected by a CCD camera when biased close to breakdown, supporting the finding of non-equilibrium in their constriction.

Next we considered the spatial variation of infrared emission and, hence, the temperature distribution along the biased graphene sample (Fig. 3). Graphene has a high room-temperature thermal conductivity, $\kappa_{Gr} = 5,000$ Wm$^{-1}$K$^{-1}$,[3, 4] therefore part of the heat is carried laterally into the metallic contacts. However, Umklapp scattering reduces this value substantially at elevated temperatures,[15-17] while surface polar phonon scattering enhances the energy transfer to the SiO$_2$ substrate. [7, 18-21] For devices that are longer than a few microns as in our case, heat transfer into the substrate dominates:[7]

$$T(x) = T_{sub} + p(x)/g. \quad (3)$$

Here $T_{sub} = 293$ K is the substrate temperature, $p(x)$ is the locally generated power, and $g$ is the effective thermal conductivity of the substrate which depends on the thermal coupling between graphene and the SiO$_2$. If $p(x)$ were position independent, we would expect a flat temperature distribution throughout the device. Temperature drops should be limited to the immediate vicinity of the contact areas. In contrast, we observe in Figure 3a that the infrared radiation is strongly position- and also gate-voltage dependent, suggesting that the temperature distribution is dominated by a spatially varying resistance, rather than the heat dissipation away from the graphene ribbon.

Figure 3 shows a broad emission maximum close to the drain contact for a gate voltage $V_g$ of about 20V that moves into the channel for higher gate voltages and is close to the source at $V_g$ around 40V. If the infrared radiation is indeed associated with the electrical power dissipated locally,

$$p(x) = I \cdot dV(x)/dx, \quad (4)$$

where $dV(x)/dx$ is the local electric field, it should be maximized at positions where the carrier density $\eta(x) = n_e(x) + n_h(x)$ is lowest. This follows directly from the requirement of current continuity,

$$I = -\mu \cdot \eta(x) \frac{dV(x)}{dx} = \text{const.}, \quad (5)$$

where $\mu$ is the carrier mobility. In general, $\mu$ depends on charge density, temperature, and electric field, but these dependencies are weak and the resulting variations are small such that to a first order, we can treat $\mu$ as constant. In graphene, the carrier density is minimized when the Fermi level, which defines the charge density $\rho(x) = n_h(x) - n_e(x)$ controlled by the gate, crosses the Dirac or neutrality point, where $n_e=n_h$. *The infrared images therefore provide to a good measure the spatial location of the Dirac point within the biased device.*

To answer why the hot zone moves between source and drain contacts upon changing the gate voltage, we note that the large source-drain bias leads to a significant potential drop along the device and thus the Fermi-level becomes position dependent. The potential within the graphene sheet can be written in terms of gate-voltage and local charge density $\rho(x) = n_h(x) - n_e(x)$:

$$V(x) = V_g - V_{Dirac} + C^{-1}\rho(x), \quad (6)$$

where $C \approx 1.15 \times 10^{-4}$ F/m$^2$ is the graphene capacitance and $V_{\text{Dirac}}$ is the Dirac voltage, which depends on the doping level in graphene and the substrate. [22] When the local potential equals $V_g - V_{\text{Dirac}}$, the net charge is zero, and the Fermi level lines up with the Dirac point. Close to the source contact, the potential is independent of gate voltage and given by $V_s - V_C$, where $V_C$ is the voltage drop at the contact. As a result, the Fermi level aligns with the Dirac point at the source contact when $V_g - V_{\text{Dirac}} = V_s - V_C$. At the drain contact, the analogous expression is $V_g - V_{\text{Dirac}} = V_d + V_C$, while halfway between the source and drain contacts, the Fermi level crosses the Dirac point at $V_g - V_{\text{Dirac}} = (V_d + V_s)/2$.

According to this simple model, we can extract the Dirac voltage from the half-way position at $V_{g(x=0)} = 32$ V (Fig. 3a): $V_{\text{Dirac}} = V_{g(x=0)} - (V_d + V_s)/2 = 47$ V. The high positive value of the Dirac voltage is likely due to trapped charges in the gate oxide or due to adsorbates on the graphene surface. Note also that $V_{g(x=0)} = 32$ V coincides roughly with the minimum in the high-bias $I$-$V_g$ characteristics around $V_g = 35$ V, as it should. From the gate-voltage range over which the hot zone moves from source to drain, we estimate a contact resistance of $2 \cdot V_C = 6$ V, consistent with the differences typically observed between two- and four-probe measurements.

In Fig. 4b we show the spatial temperature profiles extracted from the infrared intensity cross sections in Figure 3c. Based on these temperatures, we obtain the potential drop along the channel by using Eqs. (3) and (4). Both electron and hole carrier densities in the channel can be extracted using Eqs. (5) and (6). The results are shown in Figs. 4c and 4d. Note that a large minority carrier population still exists tens of microns away from the position of the Dirac point. The Dirac point, located where $n_e = n_h$, moves from drain to source by changing the gate voltage (Fig. 4e). The rate of the neutrality point motion $dX_{\text{Dirac}}/dV_g$ to a first order is given by the inverse of the source-drain electric field $\approx L_{\text{ch}}/(V_{sd} - 2 \cdot V_C)$, where $L_{\text{ch}}$=55 µm is the graphene channel length.

Aside from the broadly peaking temperature profile that is associated with the position of the Dirac point in the graphene device, we often observe stationary hot spots that do not move with changing gate voltage. These spots are likely associated with defects. Occasionally, we also saw local hot spots that reversibly turn into cool spots when the carrier type is inverted by changing the gate voltage. One such spot can be seen in Fig. 3 at about 2 µm from the center of the channel (arrows). We explain the existence of the spot as being the result of locally trapped charges in the oxide that produce image charges in the graphene. In this case the trapped charges should be holes, because the measured temperatures are consistent with a decreased local charge density under p-type conduction and an increased local charge density under n-type conduction.

In conclusion, we investigated the thermal radiation from graphene that is self-heated by an electrical current. We found a wavelength independent emissivity of $\varepsilon=(1.6\pm0.8)\%$ in the near-infrared, in agreement with measurements of optical absorption. Bias voltages determine not only the absolute temperature, but also the temperature profile in large-scale graphene. In particular, the temperature maximum along the graphene sample reveals the location of the conductivity minimum at the Dirac point.

**Methods**

Graphene was exfoliated from graphite. Large, rectangular graphene flakes were defined by e-beam lithography and oxygen plasma etching. The Ti/Pd/Au electrodes were also lithographically defined. The silicon substrate, separated by 300 nm $SiO_2$ from the graphene was used as the backgate. Devices are intentionally large (for example 55 μm long and 3.5 μm wide), so that infrared emission from the graphene can be spatially resolved in the far-field. At an emission wavelength of 2 μm we estimate a spatial resolution of the setup on the order of 3 μm. Single-layer graphene devices were identified by the green-light method, [23] and confirmed with Raman spectroscopy. For Fig. 2, a graphene sample with a lithographically-defined constriction (length 4.15μm, width 1.45μm) was used that selectively heated up at that position during electrical transport.

The optical measurements were done in a vacuum chamber ($P\sim10^{-5}$ Torr), using a 20x NIR objective with long working distance in front of a transparent window. The entire setup is shown in Figure 1a. Devices were kept in vacuum for several days before the measurements in order to render them ambipolar (in air they behaved as p-type). Various cooled short-pass filters in the range of 1800 nm to 2500 nm were used to reduce the dark counts of the LN-cooled HgCdTe detector array. A transmission grating fabricated on top of a prism was used to disperse the infrared radiation perpendicular to the direction of the graphene strip. The grating/prism is fabricated so that the 1st order of the diffracted light around 1600 nm follows the normal direction. This makes it very convenient to switch between spectroscopy and imaging modes simply by inserting or removing the grating/prism.

A second, similar setup was used to measure the devices in air with a 60x 0.7NA objective. The measurements for Fig. 2 were performed on this setup, because it allowed measuring the thermal radiation and the Raman spectrum at the same time. This setup could also be calibrated with a commercial blackbody source of known temperature and emissivity ($\varepsilon=0.95$). All emissivity measurements were done on this setup in the unipolar current carrying regime, using a set of narrow-band filters instead of the grating/prism combination. This was necessary because the blackbody source is an extended object and the spectroscopy with the first setup only works if the emission originates from point or line sources.

After the thermal measurements were performed, the single-layer graphene device was cut to fabricate Hall-bars in order to measure its four probe low-field mobility:

$\mu_0$=2400 cm$^2$/Vs. At high bias, the mobility scales as $\mu = \mu_0/(1+\mu_0 \cdot F/v_{sat})$, where F is the electric field and $v_{sat}$ is the saturation velocity. The calculated high-bias mobility $\mu$=1860 cm$^2$/Vs was used in eq. 4 to extract the carrier densities in Figs. 4c and d.

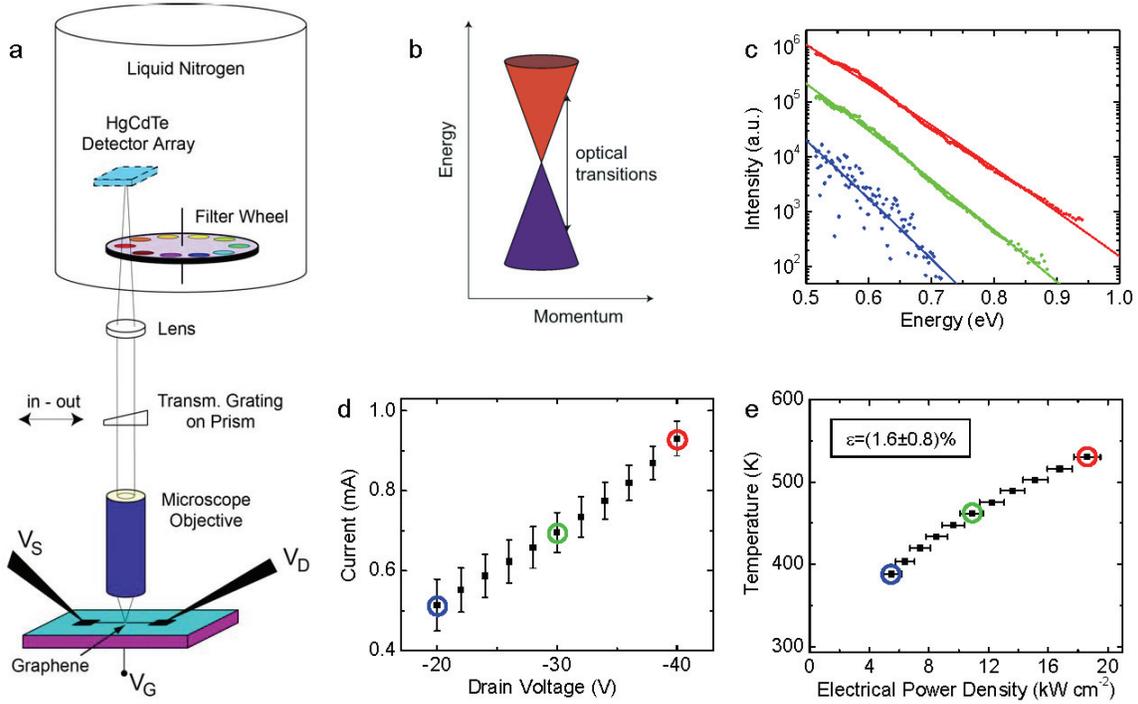

**Figure 1: Thermal emission from biased graphene. a,** Experimental setup, consisting of a vacuum probe station, microscope, and infrared imaging detector. A transmission grating is inserted for spectroscopy and removed for imaging, while a set of short-pass filters is used to cut down on the detector dark counts. **b,** Sketch of the graphene electron dispersion in the vicinity of a Dirac point. **c,** Measured infrared radiation from biased graphene for three different electrical power levels (5.5 kW/cm$^2$ (blue), 10.9 kW/cm$^2$ (green), 18.6 kW/cm$^2$ (red)). Solid lines are fits to eq. 1 (radiation of a grey body). **d,** Corresponding $I$-$V_d$ characteristic ($V_g$=20 V). The error bars are due to slight hysteresis during the measurement. The values corresponding to the three curves shown in panel 1b are highlighted in blue, green, and red. **e,** Temperature as a function of electrical power, extracted from the grey body fits in panel 1c. The measured emissivity of the single-layer graphene sample is indicated.

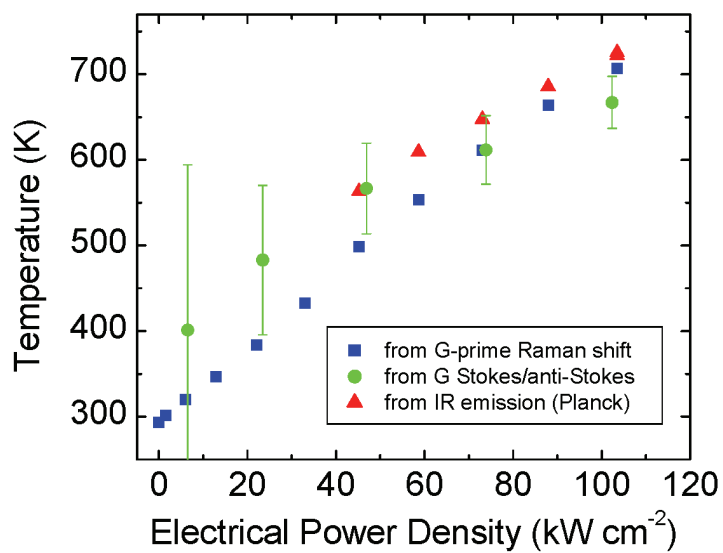

**Figure 2: Comparison of three different temperature measurements.** The temperatures extracted from blackbody fits (red ▲), are compared to the temperatures extracted from the G-prime phonon softening (blue ■) and the Stokes/anti-Stokes ratio of the G-band intensity (green ●). These measurements were done on a different sample with a 4.15μm by 1.45μm constriction that acted as a local hot spot. The blackbody fit agrees very well with the two established methods. Statistical error bars for the G-prime shift and blackbody fit are within the symbols.

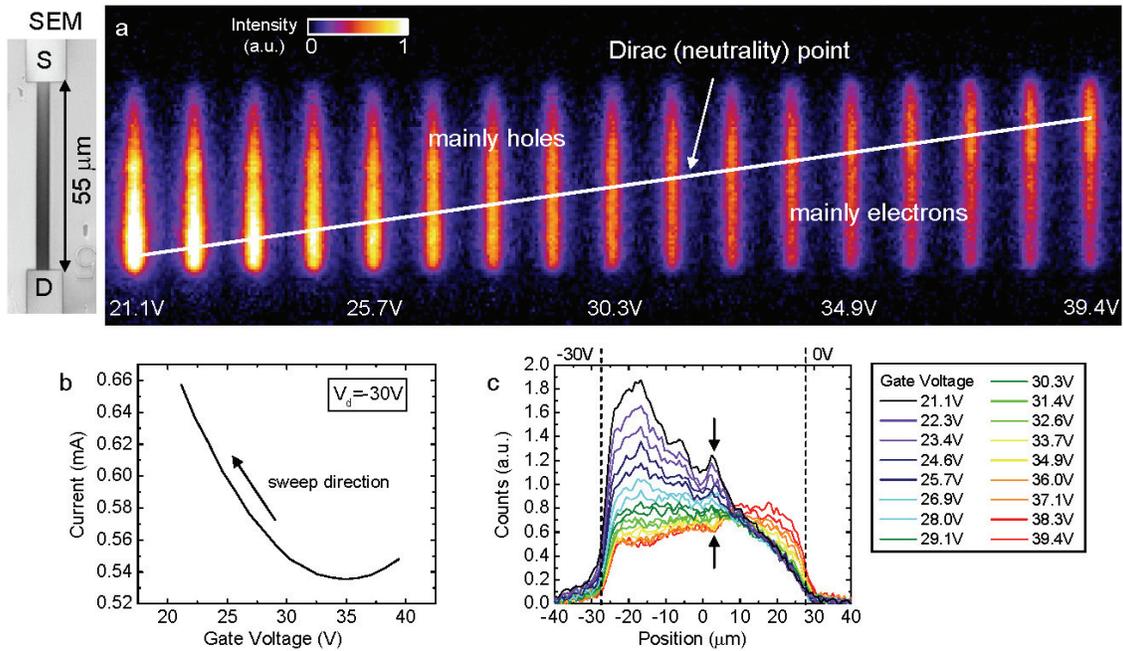

**Figure 3: Bias-dependent thermal images of graphene. a,** Spatial images of the integrated infrared emission (with wavelength up to 2000nm) from the graphene sample from Fig. 1. The drain bias was $V_d$=-30 V and the gate voltages varied between $V_g$=20 V and 40 V as indicated. The SEM image shows the graphene contacted by source (S) and drain (D) contacts. The graphene becomes hottest at the position of the Dirac point, which can be moved by the gate voltage. The white line is a guide to the eye. **b,** Corresponding *I-$V_g$* characteristic. **c,** Infrared intensity profile along the length of the graphene sample, extracted from the images in Fig. 3a. The dashed lines mark the position of drain and source contacts. The arrows point to a local hot spot under hole conduction that reversibly turns into a cold spot under electron conduction.

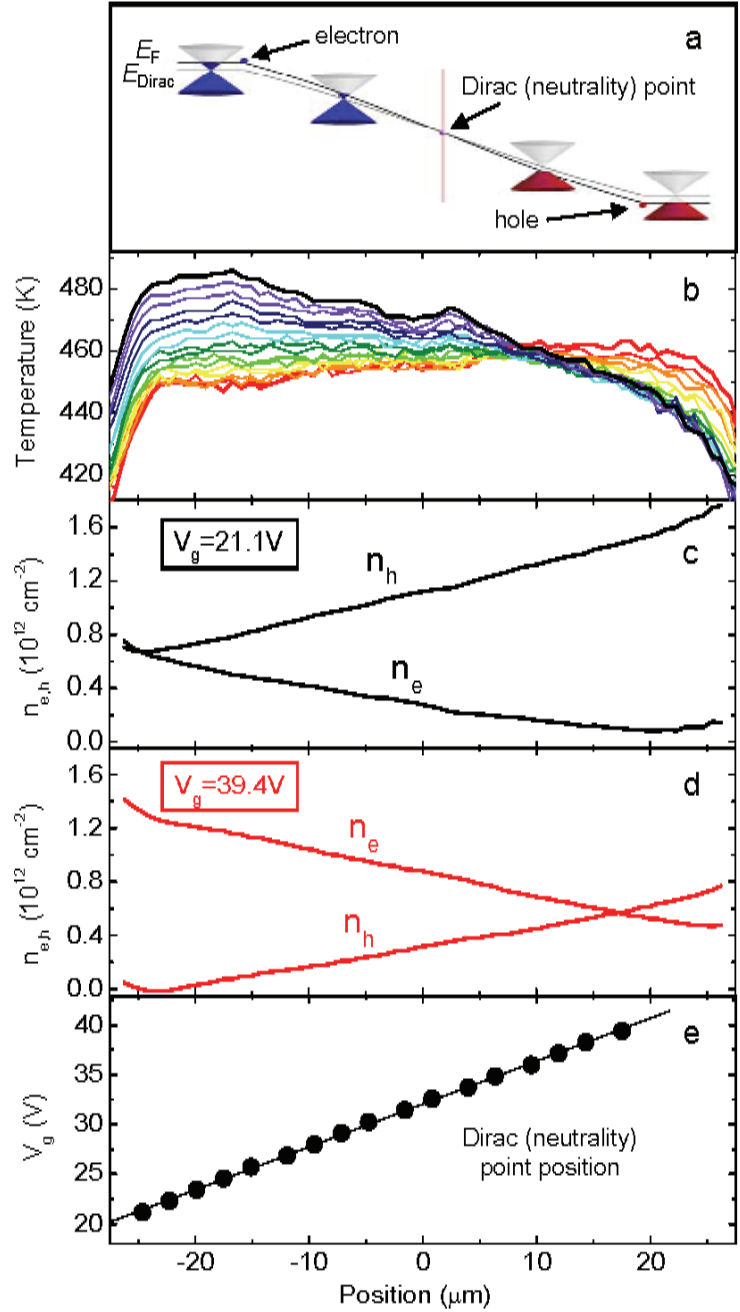

**Figure 4: Temperatures, charge carriers, and the position of the Dirac point during gate-voltage sweeps. a,** Schematic of the Fermi-level within the graphene device in the ambipolar regime. **b,** Temperature profile along the graphene sample for the different gate voltages, extracted from the infrared intensity images in Fig. 3c, black curve – $V_g$=21.1 V, other curves are obtained with a step of $\delta V_g$=1.14 V, red curve – $V_g$=39.4 V. Electron and hole distribution in the device for **c,** $V_g$=21.1 V and **d,** $V_g$=39.4 V using measured temperatures and Eq. (3-6) with high-field mobility $\mu$=1860 cm$^2$/Vs and contact resistance $2\cdot V_C$ = 6 V. **e,** Dirac (neutrality) point position vs. gate voltage. The solid line is a linear fit.